# A new bibliometric approach to assess the scientific specialization of regions[1]

**Abstract**

The objective of the current work is to identify the territorial scientific specializations present in Italy, at the levels of regions and provinces (NUTS2 and NUTS3). To do this, we take a bibliometric approach based on the scientific production of the entire public research system in the hard sciences sphere, for the five years 2006-2010. In particular, we apply a new index of scientific specialization (Scientific Specialization Index, SSI) that takes account of both the quantity and quality of scientific production achieved by the research institutions of a given territory.

# 1. Introduction

A regional innovation system is conceived as involving various organizations concentrated in a geographical area - such as universities, public research institutions, companies and agencies active in technological transfer - which create, disseminate and apply new knowledge through interactive, cooperative activity. In recent decades, the "region" has gradually become recognized as the territorial area suitable for strategic action in the development of innovation-based learning economies. Associated with this, the concept of the regional innovation system has become a major subject area in economic literature (Doloreux and Parto, 2005). Such literature stresses the role of the endogenous approach to local and regional development policy, based on the idea that regional development and the resulting economic growth are driven by the endogenous forces of knowledge and technology developed in the region (Tödtling, 2010; Asheim and Isaksen, 1997; Foray and Lundvall, 1996). Geographic concentration of research and development activity (R&D) may provide competitive advantages for a given territory, through potential spillovers made possible by closeness of R&D actors and new knowledge users, clustered in specific areas (Matthiessen et al., 2002).

Until recently, policy makers have primarily been interested in understanding the relative technological specializations of regions, or in other words the comparative advantages that nations or territories might enjoy in the technological dimension, compared to their counterparts. The data concerning technological specialization provide a basis for analysis of a territory's relative competitiveness and the study of its innovation potential and technology spillovers (Kumar and Siddharthan, 1997). The corresponding information on scientific specializations has usually been considered less important, in the belief that scientific activity does not impact in a direct way on a territory's economic performance, at least not in the short to medium term. However in recent decades, the time necessary to incorporate scientific discoveries in technological innovation has progressively diminished. At the same time, there have been new demands for a change in mission among universities and research institutions, away from the simple role of "knowledge mill", imparting know-how (skills and capabilities) and "know-why" (theories, principles), towards new roles as protagonists in the development of their local territories. Current trends thus include an increased focus on exploiting research results, and generally towards greater private-sector partnerships (Charles, 2006; Kitagawa, 2004; Thanki, 1999; Garlick, 1998).

For policy makers, who are increasingly focused on regional-level public R&D investment, there is also obviously a new emphasis on access to information concerning the differences in regional scientific capacities, which can support the identification of priorities and the efficient allocation of the territorial funding (Peter and Frietsch, 2009).

In the literature, the characterization of the scientific profile of a given territory is typically conducted by gathering and analyzing bibliometric data: specifically by analyzing the geographic distribution of scientific production, as indexed in the major bibliometric databases. Frenken et al. (2009) offer a particularly useful review of the full range of scientometric studies that analyze the spatial dimension of scientific production. Most of these studies are based on observation of national data (May, 1997; Adams, 1998; Cole and Phelan, 1999; Glänzel et al., 2002; King, 2004; Leydesdorff and Zhou, 2005; Horta and Veloso, 2007). Analyses at the regional level have been less frequent: one case is the work by Matthiessen and Schwarz (1999), on the analysis of aggregated publication records for European metropolitan areas for the years 1994-1996. A second large-scale analysis at the regional level was conducted by Acosta et al., (2012): here the aim was to identify the spatial distribution of academic scientific production across European regions for the period 1998-2004, thus providing policy makers with mapping and information on scientific activity in the EC European Research Area.

Within Italy, Tuzi (2005) pioneered bibliometric measures of the scientific specialization of regions, by two separate indicators: one based on publications and the other on average citations per paper. Morettini et al. (2012) document "knowledge activities" at the regional level, through the measurement of R&D expenditures, patents, and publications originating from "local labor systems". Again regarding Italy, Abramo et al. (2013) have recently offered a spatial analysis of the new knowledge supply from Italian public research institutions. This contribution is held to have broader significance, most notably for the methodological approach. The authors use citations, and not simply the counting of publications, to map the territorial distribution of new knowledge: in fact, counts of publications alone do not permit an assessment of the real value of the new knowledge produced.

Continuing from their preceding work, these same authors now further elaborate their methodology, and apply it for the purposes of analyzing the scientific specialization of Italian regions and provinces. The analysis simultaneously reveals: i) for each territory, the scientific specializations that are present and ii) for each scientific field, which are the territories most specialized in that field. Findings of this type can inform public research and industrial policies at the national and regional levels, as well as the localization strategies of hi-tech companies.

The analysis is based on bibliometric data from the Thomson Reuters Web of Science (WoS). Beginning from the publications indexed between 2006 and 2010 and produced by researchers on staff at Italian public research organizations, the indicator "Scientific Specialization Index" is calculated, based on the standardized citations received by these publications, in a manner taking due account of both the quantity and the impact of the scientific production from the public research organizations situated in a given territory.

Given the study objective as described above, the next section of the study presents the methodological aspects of the analysis: field of observation, dataset, and sources. Section "Results and analysis" presents and discusses the results of the analysis and section "Conclusions" concludes with the authors' comments.

## 2. Methodology

To relate the territorial framework for our study to the broader context of European statistics and analysis, we refer to the European Union Nomenclature of Territorial Units for Statistics (NUTS)[2]. Specifically, the legislative context in Italy provides for the administrative and territorial subdivisions known as "regions" (NUTS 2) and "provinces" (NUTS 3). There are 20 regions and, during the period under observation, these were further subdivided in a total of 110 provinces. The scientific production of public research institutions is extracted from the Italian Observatory of Public Research (ORP)[3], a database developed and maintained by the authors and derived under license from the Thomson Reuters WoS. Beginning from the raw data of the WoS and applying a complex algorithm for reconciliation of bibliometric addresses, each publication is attributed to the organizations of its co-authors, and consequently to the territory where they work. The algorithm is based on a controlled vocabulary of over 30,000 rules (D'Angelo et al., 2011) [4]. Unlike the arts and humanities and some fields of the social sciences, in the hard sciences the prevalent form of codification for research outputs is publication in scientific journals. Other forms of output are often followed by publications that describe their content in the scientific arena. Thus analysis of publications alone permits derivation of mapping that is certainly representative of the new knowledge produced by public research organizations, providing that the field of observation is limited to the subject categories (SCs) of the hard sciences[5] (a total of 167 SCs, according to WoS classification[6]).

---
[2] NUTS is a geocode standard for referencing the subdivisions of countries for statistical purposes. The standard is developed and regulated by the European Union, and thus only covers the member states of the EU in detail.

The data extracted thus concern the scientific production achieved in the given subject categories over the 2006-2010 period, by all national public research organizations, meaning all Italian universities (95), research institutions (76) and research hospitals (200). This dataset of the 2006-2010 Italian scientific production (articles, reviews, proceeding papers, letters) in the hard sciences consists of roughly 260,000 publications, authored by public research organizations located in 101 out of the total 110 provinces.

To assess the relative public supply of knowledge at the territorial level we do not simply count the publications produced, but rather consider their real value in terms of impact on the advancement of knowledge. As proxy of value, bibliometricians adopt the number of citations received by the publication. Because this number is a function of the time elapsed from the publication date, as well as of the SC of the publication, we need to standardize the citations. To that end, we use a relative indicator, named Article Impact Index (AII), given by the ratio of the number of citations received by a publication (as of 31/12/2011) to the average of the citations for all the other national publications of the same year and WoS subject category[7] (Abramo et al., 2012). For each subject category (SC), the values of AII are successively aggregated at the provincial level (NUTS3) and then the higher level of the region (NUTS2) to obtain an indicator named Scientific Strength (SS), given by the sum of the Impact Index (AII) of all the publications produced in the particular territory. Any publications co-authored by scientists working in organizations of the same territory are counted only once for that territory. In assigning a publication to a territory we do not adopt fractional counting in function of the number of authors. The reasoning for these last-described procedures is that a publication represents new knowledge produced in a territory independently of the number of people in that territory that contributed to its production. For publications in multi-category journals, we attribute each SC a fractional value of AII, equal to the inverse of the number of subject categories included in the journal.

---

[3] www.orp.researchvalue.it. Last accessed November 19, 2013.
[4] As an example, the rules resolve 142 variants of "University of Rome 'Tor Vergata'", detected in WoS bibliometric affiliations for the period under examination.
[5] Biology, Biomedical research, Chemistry, Clinical medicine, Earth and space sciences, Engineering, Mathematics, Physics.
[6] http://ip-science.thomsonreuters.com/cgi-bin/jrnlst/jlsubcatg.cgi?PC=K, last accessed November 19, 2013
[7] The subject category of a publication corresponds to that of the journal where it is published. For publications in multidisciplinary journals the scaling factor is calculated as a weighted average of the standardized values for each subject category.

To determine the scientific specialization of territories we use an indicator named Scientific Specialization Index (SSI). In operational terms, SSI is calculated applying the Revealed Comparative Advantage (RCA) methodology and, in particular, the *Balassa index* (Balassa, 1979). The Scientific Specialization Index of the territory *k* in the SC *j* ($SSI_{kj}$) is therefore defined as:

$$SSI_{kj} = 100 * \tanh ln \left\{ \frac{(SS_{kj}/\sum_i SS_{ki})}{\sum_k SS_{kj}/\sum_k \sum_i SS_{ki}} \right\} \qquad (1)$$

with $SS_{ki}$ indicating la Scientific Strength of the territory *k* in the SC *i*. Use of the logarithmic function centers the data around zero and the hyperbolic tangent multiplied by 100 limits the $SSI_{kj}$ values to a range of +100 to -100. The closer the value of the index is to +100, the greater is the specialization of the territory in the SC and, vice versa, the closer the index approaches to -100, the less the territory is specialized in the SC. Values around zero are labeled as "expected" or "national average".

The SSI is conceptually similar to the renown Activity Index (AI) introduced by Frame (1977). The AI indicates whether a country has a relatively higher or lower share in world publications in particular fields of science than its overall share in the world total of publications. Other scholars (Schubert and Braun, 1986; Schubert et al., 1989) applied then the same indicator. A mathematical variation of the AI is the Relative Specialization Index (RSI), whose values range from -1 to +1 (REIST-2, 1997). More recently Rousseau and Yang (2012) observed some theoretical problems in the construction of the activity index (AI) and related indicators, due to the mathematical structure of this indicator. The main difference between the AI and the SSI is that the latter weights each publication by its normalized impact.

## 3. Results and analysis

We begin the analyses of scientific specialization at the higher territorial level of the regions and then proceed to the provinces. For better comprehension of the Italian territorial system, for each region Table 1 shows: the list of its provinces, the numbers of inhabitants, the number of research organizations located in the region and the WoS publications produced by researchers in these organizations, over the period under observation. The analysis at the regional level is more apt to inform research and industrial policies of the national governments, while that at the provincial level should be of interest to the regional governments.

### 3.1 Analysis at the regional level

In the first analysis of scientific specialization at the regional level we identify the dominant SCs in each region. Thus for each region, Table 2 presents the first three SCs by value of SSI: The most frequently observed SC is Engineering, petroleum, which is the top scientific specialization for the regions of Basilicata, Molise, Umbria and Campania.

There are five regions (Basilicata, Marche, Molise, Trentino-Alto Adige, Valle

D'Aosta) that show values of SSI greater than 90 for all three of the first SCs. In these cases, the SCs involved are consistently from five of the eight scientific-technological disciplines (Engineering, Clinical medicine, Mathematics, Biology, Earth and space sciences). We observe that, apart from the Marche, these cases all involve regions that are small in both land area and population, and with scientific production that represents only a few percentage points out of the national total. Also, these particular SCs are generally research niches, with very limited overall scientific production. Four regions (Campania, Emilia Romagna, Lazio, Lombardy) show values of SSI that are below 80: these are large, heavily populated regions with high concentrations of universities and public research institutions, thus responsible for a wide range of research activities. Lombardy is particularly notable for the three values of SSI all falling below 70.

[Insert Table 1 here]
[Insert Table 2 here]

We now present the companion analysis to Table 2: after first identifying the 20 SCs with overall highest values of SSI, we then list for each one of these, the three regions with the highest respective values of "territorial" SSI (Table 3).

The 20 SCs identified belong to seven of the eight hard science disciplines (only Chemistry is missing) but these are represented with differing frequency, ranging from once only for Physics and Biomedical research to six times for Biology. The 60 highest-value positions presented in Table 3 are occupied by 13 of the possible 20 regions, however these also occur with differing frequency: the minimum occurrence (1) is for Sicily and the maximum (13) is for Basilicata. At the top of the list, Operations research & management science (Mathematics) shows a very high range of variation in SSI, from the peak of 100 for Valle d'Aosta[8] to 31.0 for Sicily. Globally, the SC with the triplet of regions scoring the highest values of SSI is Forestry (Biology), related in particular to: Basilicata (98.6), Molise (98.5) and Trentino-Alto Adige (90.8). This is a combination of a very specialized scientific field and small-sized regions; and certainly in the case of Basilicata and Molise they are regions near the bottom for share of national scientific production. Returning to an overall view, Table 3 also shows Basilicata scoring in the top three values of SSI for a full 13 of the 60 SCs, followed by Valle D'Aosta in 8, and Molise (7), the Marche and Sardinia (6). In contrast, the three very populous regions (Lombardy, Campania and Lazio) do not appear at all.

[Insert Table 3 here]

For any given SC, presentation by radar diagrams offers the best possibility of visualizing the regional distribution of specialization. Figure 1 presents, for the 20 Italian regions, the SSI values for eight SCs (i.e. an example for each discipline, chosen randomly): Biochemistry & molecular biology; Cardiac & cardiovascular systems; Pharmacology & pharmacy; Chemistry, organic; Geochemistry & geophysics; Astronomy & astrophysics; Computer science, theory & methods; Mathematics, applied. The reference line is at zero, which would represent agreement with the national average and thus the absence of specialization. In Biochemistry & molecular biology there is a relatively uniform distribution of regional SSI values. Lombardy, Puglia, Tuscany and Veneto show values very close to zero, indicating production of new knowledge at levels very close to the national average for this SC. In contrast, Campania and Molise show values greater than 30, indicating a certain level of scientific specialization; Basilicata and Trentino-Alto Adige are in the opposite situation of very negative values of SSI, thus indicating high de-specialization in this area. The Computer science, theory & methods SC presents a more polarized distribution of specialization: Abruzzo and Campania show values near zero (less than 10), indicating a very near absence of specialization; however Calabria, Trentino-Alto Adige and Valle d'Aosta show very high positive values (greater than 80), indicating a notable level of specialization. At the opposite extreme, still for the same SC, seven regions (Basilicata, Friuli Veneto Giulia, Lazio, Lombardy, Marche, Molise and Sicily) result as strongly non-specialized, with very negative values of SSI (less than -50).

---

[8] This region represents an outlier: it is the smallest of the 20 Italian regions, with only 126,000 inhabitants and three research organizations situated in the territory, producing a total of 30 publications in the period under analysis.

*Figure 1. Regional distribution of SSI for eight subject categories (one example for each discipline analyzed)*
[Insert Figure 1 here]

### 3.2 Analysis at the provincial level

We now conduct the analysis at the more detailed territorial level of the Italian provinces. For each province we measure the index of specialization in every active SC. For reasons of space, we present only the cases of the "capital city" provinces of the twenty regions: Table 4 indicates the top three SCs in the scientific specialization for these provinces.

The 60 positions listed in the table are occupied by 46 of the possible 167 SCs: Engineering, petroleum is the most frequent SC, occurring in 5 cases. The 60 values of SSI vary from a minimum of 65.0 for the province of Milan and the SC of Oncology (Biomedical research) to a maximum value of 100 for the province of Aosta in Operations research & management science (Mathematics). As we would expect, these two extreme values refer to the among the most populous (Milan) and least (Aosta) of all Italian provinces. Milan actually presents all three of the lowest overall values of SSI: Ornithology (69.4), Cell & Tissue Engineering (65.4) and Oncology (65.0). Examining in further detail, we observe that of the 60 values of SSI, a full 21 concern the Engineering discipline and 17 fall under Clinical medicine; while there is only one SC each for Chemistry and Mathematics.

[Insert Table 4 here]

Continuing the analysis at the provincial level, and similar to Table 2 for the regions, Table 5 shows the three provinces that are most specialized in the top 20 SCs for value of SSI. The 20 SCs belong to six of the eight disciplines of the hard science disciplines (Chemistry and Physics are missing). The 60 positions of the table are occupied by 36 provinces out of the total 101, almost equally divided between north, central and southern Italy. The maximum frequency is three occurrences, seen for six provinces of medium-small size (Gorizia, Matera, Savona, Taranto, Trapani, Vercelli). As we would expect, the maximum value of SSI is consistently equal to 100; the minimum observation (86.1) occurs for the province of Pesaro-Urbino, relative to the Mineralogy

SC. A full 55 out of the 60 SSIs exceed 95.

[Insert Table 5 here]

Next we wish to identify the provinces where the levels of scientific specialization in the research conducted are most extreme. To do this, we begin by considering only those province-subject category pairs where the corresponding values of SSI are greater than 50 or less than -50: in taking this step we are first isolating the pairs that indicate situations of either strong scientific specializations in the territory or the contrary situation of strong de-specialization. For each province we then calculate the ratio between the numbers of such extreme SCs and the number of SCs in which the province is active. Table 6 presents the first 20 provinces on the basis of decreasing value of the first ratio, in which the numerator is the number of highly-specialized SCs.

The first observation to make is that the provinces listed are all of medium-small size (in terms of population) and that none of them are the capital provinces for their region. The large part of these provinces also have no main university campus (although some have satellite campuses), and in general the scientific production achieved in the territory is marginal relative to the national total. The fact of low scientific production is also demonstrated by the small numbers of active SCs: these are consistently below 50, meaning that less than 30% of the 167 SCs are present. The ratio of highly-specialized SCs to active SCs is always greater than 50%, with peaks of 100% for a full eight provinces.

[Insert Table 6 here]

Next we present a complementary analysis to the preceding one: Table 7 shows the list of the first 20 SCs by incidence of the ratio of the highly-specialized provinces in those SCs to the number of provinces with publications in the given SC. For the most part, the SCs involved refer to niche scientific sectors where there is numerically limited scientific production. The number of provinces active in these SCs varies from a minimum of 12, for Engineering, marine to a maximum of 77 for Hematology. The ratio of highly specialized provinces/active provinces varies from a minimum of 0.32 for the SCs of Hematology, Food science & technology and Marine and freshwater biology, to a maximum of 0.47 for Engineering, petroleum. These results can be explained considering that the Engineering, petroleum SC has very little activity in Italian territory (being present in only 17 provinces), when compared to SCs such as Hematology (present in 78 provinces). On the other hand, the ratio of non-specialized provinces/provinces active varies from a minimum of 0.15 for the Ornithology SC to a maximum of 0.49 in Oceanography.

[Insert Table 7 here]

Finally we examine the combined analyses of the territorial distribution of public-sphere production of new knowledge, as mapped using three indicators. Other than SSI we consider the absolute value of scientific strength, and thirdly its ratio per inhabitant of the provinces. Figure 2 illustrates the mapping for data on the Italian provinces, for the example of the Biochemistry & molecular biology SC. We observe that the 10 provinces with the highest absolute values of SS are Bologna, Florence, Genoa, Milan,

Naples, Padua, Pavia, Rome, Turin and Trieste. However, factoring in the population of the provinces, only the five smallest provinces of this list remain in the first 10 positions, in the classification for SS per inhabitant (Florence, Genoa, Padua, Pavia, Trieste). Further, considering the value of SSI registered for these five provinces, we observe that all of them place below 10$^{th}$ position in national rank. Thus the more populated provinces, where there are higher number of research organizations (typically medium to large size), therefore having ample and diversified scientific production, top the rankings for SS, but then fall back many positions in SS per inhabitant and also in SSI. In contrast, the highest values of SSI are observed in little-populated provinces that host very small numbers of research organizations (primarily special-focus). The comparison between the three types of map thus permits a differentiated representation of the territorial distribution of new knowledge production, which responds to the different information needs of the policy makers. While the map based on SS depicts the distribution of the mass of new knowledge produced by the public system in a given research field, the map where this mass is related to the number of inhabitants permits the policy-maker to take due account of the macro-economic character of the territory. Finally, the map based on SSI provides a third indication concerning the extent that research and results achieved in a given field are a specific feature of a given territory (representing its comparative advantage), with respect to the national average.

*Figure 2. Provincial distribution of SS, SS per inhabitant and SSI for the subject category Biochemistry & molecular biology*
[Insert Figure 2 here]

### 4. Conclusions

A territory (nation, region or province) generally demonstrates a distinctive scientific profile, remaining quite stable over time, as a direct consequence of policy makers' selected priorities for disciplinary expenditures in R&D (Peter and Frietsch, 2009). In the current work we have applied a new bibliometric approach to measure scientific specialization at the regional and provincial levels of the Italian public research system. In place of the simple counting of publications, the method applied begins from the citations to the works, which are much more reliable in measuring the real impact of the new knowledge produced. In operational terms we use the bibliometric indicator of Scientific Strength (SS), which accounts for both the quantity and the impact of scientific production, and then calculate an index of specialization, named Scientific Specialization Index (SSI).

We first note that high values of SSI for a particular territory can be the fruit of marginal overall scientific production in quantitative terms, relative to the national total. Such effects occur when the territory has very few, small research organizations, which are of limited scope of research fields.

In effect, the analyses conducted show that at the regional level, the lowest values of SSI are obtained for the most populated regions, where there are a relatively high number of large-sized research organizations, active over a wide spectrum of hard sciences subject categories. In contrast, the highest values of SSI are detected for low-population regions where there are few research organizations, which are primarily special-focus. The phenomena observed at the regional level appear further magnified at the level of the provinces. In fact the first positions for value of SSI are occupied by small provinces, both in terms of their surface area and numbers of inhabitants, with scientific production that represents only a few percentage points out of the national totals, in part because of the absence of universities and large research institutes. In these provinces there are typically only a few SCs present, in which high specialization is possible precisely because of the low quantity and concentration of overall production. In contrast, three provinces (Milan, Naples and Rome) show the lowest values of positive SSI: these are large and heavily populated territories, with a high concentration of universities, public research institutions and specialized research hospitals. Consequently these achieve a significant percentage share of national scientific production, with contributions of differing intensity throughout the wide range of disciplines and SCs that they cover.

Analyses of this type are useful to the national and local public decision maker, for formulation of research and industrial policies, but also to managers of high-tech companies, for informing the choice of locations for R&D activities.

**References**

Abramo, G., D'Angelo, C.A., Di Costa, F. (2013) A spatial analysis of the new knowledge supply of the Italian public research institutions. *Working paper LabRTT*, http://www.disp.uniroma2.it/laboratoriortt/TESTI/Working%20paper/Spatial_analysis%20.pdf (last accessed November 19, 2013).

Abramo, G., Cicero, T., D'Angelo, C.A., (2012) How important is choice of the scaling factor in standardizing citations? *Journal of Informetrics*, 6(4), 645-654.

Acosta, M., Coronado, D., Ferrándiza, E., Leóna, M.D. (2012) Regional Scientific Production and Specialization in Europe: The Role of HERD, *European Planning Studies*. doi: 10.1080/09654313.2012.752439

Adams, J. (1998) Benchmarking international research, *Nature*, 396, 615-618.

Asheim, B., Isaksen, A. (1997) Location, Agglomeration and Innovation: Towards Regional Innovation Systems in Norway? *European Planning Studies*, 5(3), 299-330.

Balassa, B. (1979) The Changing Pattern of Comparative Advantage in Manufactured Goods, *Review of Economics and Statistics*, 61(2), 259-266.

Charles, D. (2006) Universities as key knowledge infrastructures in regional innovation systems, *Innovation*: *The European Journal of Social Science Research*, 19(1), 117 - 130.

Cole, S., Phelan, T.J. (1999) The scientific productivity of nations, *Minerva*, 37(1), 1-23.

D'Angelo, C.A., Giuffrida, C., Abramo, G. (2011) A heuristic approach to author name disambiguation in large-scale bibliometric databases, *Journal of the American Society for Information Science and Technology*, 62(2), 257-269.

Doloreux, D., Parto, S. (2005) Regional innovation systems: Current discourse and unresolved issues, *Technology in Society,* 27(2), 133-153.

Foray, D., Lundvall, B. (1996) The knowledge-based economy: from the economics of knowledge to the learning economy, in*: OECD (eds.), Employment and Growth in the Knowledge-Based Economy*, 3-28 (Paris: OECD).

Frame, J.D. (1977) Mainstream research in Latin America and the Caribbean. *Interciencia*, 2, 143-148.

Frenken, K., Hardeman, S., Hoekman, J. (2009) Spatial scientometrics: Towards a cumulative research program, *Journal of Informetrics,* 3(3), 222-232.

Garlick, S. (1998) *Creative Associations in Special Places: Enhancing the Role of Universities in Building Competitive Regional Economies*, (Canberra: Deetya).

Glänzel, W., Schubert, A., Braun, T. (2002) A relational charting approach to the world of basic research in twelve science fields at the end of the second millennium, *Scientometrics*, 55(3), 335-348.

Horta, H., Veloso, F. M. (2007) Opening the box: Comparing EU and US scientific output by scientific field, *Technological Forecasting and Social Change*, 74(8), 1334-1356.

King, D.A. (2004) The scientific impact of nations, *Nature*, 430, 311-316.

Kitagawa, F. (2004) Universities and regional advantage: higher education and innovation policies in English regions, *European Planning Studies*, 12(6), 835-852.

Kumar, N., Siddharthan, N. S. (1997) *Technology, Market Structure and Internationalization: Issues and Policies for Developing Countries*, (London and New York: Routledge and UNU Press).

Leydesdorff, L., Zhou, P. (2005) Are the contributions of China and Korea upsetting the world system of science? *Scientometrics*, 63(3), 617-630.

Matthiessen, C. W., Schwarz, A. W., Find, S. (2002) The top-level global research system, 1997-99: centres, networks and nodality. An analysis based on bibliometric indicators, *Urban Studies*, 39(5-6), 903-927.

Matthiessen, C. W., Schwarz, A. W. (1999) Scientific centres in Europe: an analysis of research strength and patterns of specialisation based on bibliometric indicators, *Urban Studies*, 36(3), 453-477.

May, R M. (1997) The scientific wealth of nations. *Science*, 275(7), 793-796.

Morettini, L. Perani, G., Sirilli, G. (2012) The concentration of knowledge activities in Italy. An analysis at local level. Paper presented at the *XXXIII Meeting of the Italian Regional Science Association,* Rome, September 13-15.

Peter, V., Frietsch, R. (2009) *Exploring regional structural and S&T specialisation: implications for policy*. Regional Key Figures of the European Research Area Booklet 2009, Fraunhofer ISI.

REIST-2 (1997) Second *European Report on Science and Technology Indicators 1997*. Second Edition, EUR 17639. European Commission, Brussels. ISBN 92-828-0271-X

Rousseau, R., Yang, L. (2012) Reflections on the activity index and related indicators. *Journal of Informetrics*, 6(3), 413-421.

Schubert, A., Braun, T. (1986) Relative indicators and relational charts for comparative assessment of publication output and citation impact. *Scientometrics*, 9(5-6), 281-291.

Schubert, A., Glänzel, W., Braun, T. (1989) World flash on basic research: Scientometric datafiles. A comprehensive set of indicators on 2649 journals and 96 countries in all major science fields and subfields, 1981-1985. *Scientometrics*, 16(1-6), 3-478.

Thanki, R. (1999) How do we know the value of higher education to regional development?, *Regional Studies*, 33(1), 84-89.

Tödtling, F. (2011) Endogenous approaches to local and regional development policy, in: A. Pike, A. Rodríguez-Pose, J. Tomaney (Eds), *Handbook of Local and Regional Development*, 333-343 (London and New York: Routledge).

Tuzi, F. (2005) The scientific specialisation of the Italian regions. *Scientometrics*, 62(1), 87-111.

*Table 1. List of Italian regions and provinces; average population and total WoS publications over the years 2006-2010*

| Macro-area | Region | Inhabitants (x 1,000) | Provinces | Organizations* | WoS Publications** | SC*** |
|---|---|---|---|---|---|---|
| Northwest | Liguria | 1,612 | Genoa; Imperia; La Spezia; Savona | 21 (2U, 9I, 10H) | 10,393 | 130 |
| Northwest | Lombardy | 9,646 | Bergamo; Brescia; Como; Cremona; Lecco; Lodi; Mantua; Milan; Monza and Brianza; Pavia; Sondrio; Varese | 74 (13U, 18I, 43H) | 53,185 | 161 |
| Northwest | Piedmont | 4,395 | Alessandria; Asti; Biella; Cuneo; Novara; Turin; Verbania; Vercelli | 29 (5U, 9I, 15H) | 19,243 | 154 |
| Northwest | Valle d'Aosta | 126 | Aosta | 3 (1U, 2I) | 30 | 1 |
| Northeast | Emilia Romagna | 4,284 | Bologna; Ferrara; Forlì-Cesena; Modena; Parma; Piacenza; Ravenna; Reggio Emilia; Rimini | 37 (5U, 12I, 20H) | 33,162 | 161 |
| Northeast | Friuli V. Giulia | 1,222 | Gorizia; Pordenone; Trieste; Udine | 23 (3U, 12I, 8H) | 12,670 | 129 |
| Northeast | Trentino A. Adige | 1,007 | Bolzano; Trento | 16 (2U, 10I, 4H) | 4,888 | 101 |
| Northeast | Veneto | 4,828 | Belluno; Padua; Rovigo; Treviso; Venice; Verona; Vicenza | 47 (5U, 13I, 29H) | 23,119 | 150 |
| Center | Lazio | 5,534 | Frosinone; Latina; Rieti; Rome; Viterbo | 52 (11U, 19I, 22H) | 47,179 | 161 |
| Center | Marche | 1,549 | Ancona; Ascoli Piceno; Fermo; Macerata; Pesaro-Urbino | 12 (4U, 5I, 3H) | 6,074 | 120 |
| Center | Tuscany | 3,675 | Arezzo; Florence; Grosseto; Livorno; Lucca; Massa Carrara; Pisa; Pistoia; Prato; Siena | 34 (7U, 13I, 14H) | 31,862 | 159 |
| Center | Umbria | 884 | Perugia; Terni | 10 (2U, 4I, 4H) | 5,539 | 107 |
| South & islands | Abruzzo | 1,323 | Chieti; L'Aquila; Pescara; Teramo | 15 (4U, 8I, 3H) | 6,763 | 112 |
| South & islands | Basilicata | 591 | Matera; Potenza | 9 (1U, 6I, 2H) | 1,638 | 48 |
| South & islands | Calabria | 2,006 | Catanzaro; Cosenza; Crotone; Reggio Calabria; Vibo Valentia | 13 (3U, 7I, 3H) | 5,878 | 108 |
| South & islands | Campania | 5,806 | Avellino; Benevento; Caserta; Naples; Salerno | 33 (6U, 15I, 12H) | 23,119 | 148 |
| South & islands | Molise | 321 | Campobasso; Isernia | 7 (2U, 5H) | 1,612 | 39 |
| South & islands | Puglia | 4,076 | Bari; Barletta-Andria-Trani; Brindisi; Foggia; Lecce; Taranto | 20 (5U, 8I, 7H) | 13,303 | 142 |
| South & islands | Sardinia | 1,665 | Cagliari; Carbonia-Iglesias; Medio-Campidano; Nuoro; Ogliastra; Olbia-Tempio; Oristano; Sassari | 14 (2U, 8I, 4H) | 6,159 | 121 |
| South & islands | Sicily | 5,029 | Agrigento; Caltanissetta; Catania; Enna; Messina; Palermo; Ragusa; Syracuse; Trapani | 30 (4U, 11I, 15H) | 18,138 | 146 |

*\* Number of research organizations located in the region (U=universities; I=public research institutions; H=Research hospitals)*
*\*\* WoS publications authored by researchers working in public organizations located in the region*
*\*\*\* Number of subject categories with at least 10 WoS publications authored by researchers of the region*

*Table 2. List of the first three subject categories (SC) by value of SSI, for each Italian region (data 2006-2010)*

| Region | SC 1 | SSI$_1$ | SC 2 | SSI$_2$ | SC 3 | SSI$_3$ |
|---|---|---|---|---|---|---|
| Abruzzo | Meteorology & Atmospheric Sciences | 92.1 | Dentistry, Oral Surgery & Medicine | 90.6 | Neuroimaging | 86.4 |
| Basilicata | Engineering, Petroleum | 99.4 | Engineering, Marine | 99.3 | Entomology | 99.1 |
| Calabria | Engineering, Manufacturing | 94.5 | Engineering, Chemical | 91.9 | Engineering, Industrial | 86.9 |
| Campania | Polymer Science | 78.5 | Soil Science | 69.1 | Engineering, Petroleum | 65.0 |
| Emilia Romagna | Materials Science, Ceramics | 76.1 | Orthopedics | 74.6 | Integrative & Complem. Medicine | 64.3 |
| Friuli V. Giulia | Engineering, Marine | 93.8 | Astronomy & Astrophysics | 85.2 | Physics, Particles & Fields | 82.5 |
| Lazio | Tropical Medicine | 73.7 | Sport Sciences | 68.7 | Parasitology | 63.3 |
| Liguria | Allergy | 96.9 | Rheumatology | 92.5 | Robotics | 88.5 |
| Lombardy | Ornithology | 65.1 | Hematology | 54.4 | Oncology | 53.1 |
| Marche | Medicine, Legal | 96.3 | Fisheries | 95.5 | Geriatrics & Gerontology | 93.9 |
| Molise | Engineering, Petroleum | 99.2 | Forestry | 98.5 | Medicine, Legal | 93.8 |
| Piedmont | Materials Science, Textiles, Paper & Wood | 93.2 | Critical Care Medicine | 79.2 | Limnology | 78.8 |
| Puglia | Rehabilitation | 84.7 | Computer Science, Cybernetics | 83.9 | Materials Science, Characterization & Testing | 82.7 |
| Sardinia | Substance Abuse | 97.9 | Agriculture, Dairy & Animal Science | 92.7 | Mycology | 82.1 |
| Sicily | Engineering, Marine | 87.7 | Physics, Nuclear | 73.5 | Chemistry, Applied | 63.2 |
| Tuscany | Andrology | 87.6 | Robotics | 71.1 | Chemistry, Medicinal | 62.3 |
| Trentino A. Adige | Remote Sensing | 96.3 | Computer Science, Cybernetics | 94.7 | Computer Science, Software Engineering | 92.5 |
| Umbria | Engineering, Petroleum | 89.2 | Mycology | 80.4 | Geriatrics & Gerontology | 74.7 |
| Valle D'Aosta | Operations Research & Management Science | 100 | Environmental Studies | 99.9 | Biodiversity Conservation | 99.7 |
| Veneto | Medical Laboratory Technology | 83.5 | Physics, Nuclear | 69.7 | Astronomy & Astrophysics | 65.6 |

*Table 3. List of the first three regions by value of SSI for the top 20 subject categories (listed by value of $SSI_1$); data 2006-2010*

| Subject Category | Discipline | Region 1 | $SSI_1$ | Region 2 | $SSI_2$ | Region 3 | $SSI_3$ |
|---|---|---|---|---|---|---|---|
| Operations Research & Management Science | Mathematics | Valle D'Aosta | 100 | Calabria | 73.8 | Sicily | 31.0 |
| Environmental Studies | Earth and Space Sciences | Valle D'Aosta | 99.9 | Sardinia | 70.1 | Basilicata | 61.2 |
| Biodiversity Conservation | Biology | Valle D'Aosta | 99.7 | Calabria | 75.9 | Sardinia | 65.8 |
| Engineering, Petroleum | Engineering | Basilicata | 99.4 | Molise | 99.2 | Umbria | 89.2 |
| Engineering, Marine | Engineering | Basilicata | 99.3 | Marche | 82.8 | Abruzzo | 79.4 |
| Mathematics, Applied | Mathematics | Valle D'Aosta | 99.2 | Calabria | 60.6 | Basilicata | 34.5 |
| Entomology | Biology | Basilicata | 99.1 | Molise | 93.7 | Trentino A. Adige | 75.9 |
| Imaging Science & Photographic Technology | Physics | Basilicata | 99.0 | Trentino A. Adige | 77.7 | Sardinia | 53.3 |
| Geography, Physical | Earth and Space Sciences | Valle D'Aosta | 99.0 | Molise | 65.3 | Basilicata | 48.7 |
| Engineering, Geological | Engineering | Basilicata | 98.9 | Campania | 64.8 | Umbria | 61.6 |
| Ecology | Biology | Valle D'Aosta | 98.9 | Trentino A. Adige | 61.9 | Sardinia | 44.8 |
| Forestry | Biology | Basilicata | 98.6 | Molise | 98.5 | Trentino A. Adige | 90.8 |
| Zoology | Biology | Valle D'Aosta | 98.5 | Sardinia | 60.5 | Marche | 31.5 |
| Substance Abuse | Clinical Medicine | Sardinia | 97.9 | Marche | 85.6 | Molise | 65.4 |
| Agronomy | Biology | Basilicata | 97.8 | Puglia | 77.3 | Molise | 70.3 |
| Meteorology & Atmospheric Sciences | Earth and Space Sciences | Basilicata | 97.5 | Valle D'Aosta | 96.9 | Abruzzo | 92.1 |
| Remote Sensing | Engineering | Basilicata | 97.4 | Trentino A. Adige | 96.3 | Campania | 41.4 |
| Allergy | Biomedical Research | Liguria | 96.9 | Campania | 52.3 | Marche | 52.3 |
| Medicine, Legal | Clinical Medicine | Marche | 96.3 | Molise | 93.8 | Puglia | 69.9 |
| Rheumatology | Clinical Medicine | Basilicata | 95.9 | Liguria | 92.5 | Marche | 89.8 |

*Table 4. List of the first three subject categories (SCs) for value of SSI, referring to each region's "capital-city" province*

| Province | SC 1 | SSI$_1$ | SC 2 | SSI$_2$ | SC 3 | SSI$_3$ |
|---|---|---|---|---|---|---|
| Ancona | Medicine, Legal | 98.5 | Fisheries | 98.2 | Geriatrics & Gerontology | 97.7 |
| Aosta | Operations Research & Management Science | 100 | Environmental Studies | 99.9 | Biodiversity Conservation | 99.7 |
| Bari | Rehabilitation | 92.8 | Parasitology | 90.0 | Computer Science, Cybernetics | 88.7 |
| Bologna | Orthopedics | 89.1 | Meteorology & Atmospheric Sciences | 82.7 | Integrative & Complementary Medicine | 79.1 |
| Cagliari | Substance Abuse | 98.5 | Psychiatry | 90.7 | Genetics & Heredity | 89.7 |
| Campobasso | Engineering, Petroleum | 99.7 | Forestry | 99.4 | Medicine, Legal | 97.5 |
| Catanzaro | Materials Science, Biomaterials | 97.9 | Engineering, Biomedical | 96.5 | Nanoscience & Nanotechnology | 94.3 |
| Florence | Andrology | 94.1 | Forestry | 83.9 | Chemistry, Medicinal | 77.8 |
| Genoa | Allergy | 96.9 | Rheumatology | 92.7 | Robotics | 88.7 |
| L'Aquila | Meteorology & Atmospheric Sciences | 98.0 | Engineering, Marine | 94.6 | Nuclear Science & Technology | 90.8 |
| Milan | Ornithology | 69.4 | Cell & Tissue Engineering | 65.4 | Oncology | 65.0 |
| Naples | Soil Science | 70.3 | Engineering, Petroleum | 70.3 | Polymer Science | 67.8 |
| Palermo | Ornithology | 92.5 | Engineering, Manufacturing | 90.9 | Engineering, Petroleum | 82.6 |
| Perugia | Engineering, Petroleum | 89.8 | Mycology | 81.4 | Geriatrics & Gerontology | 76.0 |
| Potenza | Engineering, Petroleum | 99.5 | Engineering, Marine | 99.4 | Engineering, Geological | 99.1 |
| Rome | Tropical Medicine | 75.2 | Sport Sciences | 70.4 | Parasitology | 65.0 |
| Turin | Materials Science, Textiles, Paper & Wood | 93.2 | Critical Care Medicine | 81.2 | Mycology | 72.2 |
| Trento | Remote Sensing | 96.9 | Materials Science, Ceramics | 91.9 | Computer Science, Software Engineering | 91.8 |
| Trieste | Engineering, Marine | 96.5 | Astronomy & Astrophysics | 91.4 | Physics, Particles & Fields | 89.7 |
| Venice | Oceanography | 99.2 | Engineering, Marine | 98.3 | Environmental Studies | 97.7 |

*Table 5. List of the three most specialized provinces for each of the top 20 subject categories by value of SSI*

| Subject category | Discipline | Province 1 | $SSI_1$ | Province 2 | $SSI_2$ | Province 3 | $SSI_3$ |
|---|---|---|---|---|---|---|---|
| Agricultural Engineering | Biology | Prato | 100 | Matera | 99.9 | Gorizia | 99.9 |
| Materials Science, Textiles, Paper & Wood | Engineering | Biella | 100 | Ravenna | 99.9 | Como | 97.6 |
| Horticulture | Biology | Grosseto | 100 | Ascoli Piceno | 99.9 | Imperia | 99.9 |
| Agronomy | Biology | Grosseto | 100 | Prato | 100 | Lodi | 99.9 |
| Mineralogy | Earth and Space Sciences | Sondrio | 100 | Vercelli | 98.8 | Pesaro-Urbino | 86.1 |
| Water Resources | Earth and Space Sciences | Sondrio | 100 | Savona | 99.8 | Taranto | 99.1 |
| Soil Science | Biology | Gorizia | 100 | Matera | 99.8 | Viterbo | 98.6 |
| Limnology | Earth and Space Sciences | Verbania | 100 | Savona | 98.9 | Trapani | 97.7 |
| Fisheries | Biology | Trapani | 100 | Livorno | 99.8 | Oristano | 99.6 |
| Oceanography | Earth and Space Sciences | La Spezia | 100 | Trapani | 99.9 | Oristano | 99.8 |
| Health Care Sciences & Services | Clinical Medicine | Rimini | 100 | Vercelli | 87.6 | L'Aquila | 86.3 |
| Materials Science, Ceramics | Engineering | Ravenna | 100 | Brindisi | 99.2 | Terni | 98.1 |
| Meteorology & Atmospheric Sciences | Earth and Space Sciences | Agrigento | 100 | Savona | 98.7 | L'Aquila | 98.0 |
| Entomology | Biology | Matera | 100 | Campobasso | 97.5 | Potenza | 97.3 |
| Geriatrics & Gerontology | Clinical Medicine | Ascoli-Piceno | 100 | Ancona | 97.7 | Benevento | 94.3 |
| Robotics | Engineering | Lucca | 100 | Frosinone | 99.8 | Pisa | 90.1 |
| Geology | Earth and Space Sciences | Arezzo | 100 | Pesaro-Urbino | 97.8 | La Spezia | 97.2 |
| Toxicology | Biomedical Research | Rieti | 100 | Taranto | 98.8 | Gorizia | 98.6 |
| Operations Research & Management Science | Mathematics | Aosta | 100 | Vercelli | 97.0 | Taranto | 95.5 |
| Biodiversity Conservation | Biology | Livorno | 100 | Viterbo | 99.8 | Aosta | 99.7 |

*Table 6. List of the first 20 province by value of the ratio of highly specialized SC to active SCs*

| Province | Active SCs | Highly specialized SCs | Non-specialized SCs | Ratio highly specialized SC/ active SCs | Ratio non-specialized SC/ active SCs |
|---|---|---|---|---|---|
| Agrigento | 4 | 4 | 0 | 1.00 | 0.00 |
| Ascoli Piceno | 13 | 13 | 0 | 1.00 | 0.00 |
| Caltanissetta | 5 | 5 | 0 | 1.00 | 0.00 |
| Grosseto | 2 | 2 | 0 | 1.00 | 0.00 |
| Nuoro | 4 | 4 | 0 | 1.00 | 0.00 |
| Prato | 2 | 2 | 0 | 1.00 | 0.00 |
| Rieti | 3 | 3 | 0 | 1.00 | 0.00 |
| Sondrio | 2 | 2 | 0 | 1.00 | 0.00 |
| Aosta | 17 | 15 | 0 | 0.88 | 0.00 |
| Asti | 19 | 16 | 3 | 0.84 | 0.16 |
| Arezzo | 6 | 5 | 1 | 0.83 | 0.17 |
| La Spezia | 23 | 16 | 4 | 0.70 | 0.17 |
| Imperia | 32 | 22 | 4 | 0.69 | 0.13 |
| Taranto | 35 | 24 | 3 | 0.69 | 0.09 |
| Gorizia | 19 | 13 | 3 | 0.68 | 0.16 |
| Oristano | 25 | 17 | 2 | 0.68 | 0.08 |
| Syracuse | 6 | 4 | 2 | 0.67 | 0.33 |
| Belluno | 14 | 9 | 2 | 0.64 | 0.14 |
| Savona | 37 | 23 | 10 | 0.62 | 0.27 |
| Matera | 50 | 29 | 9 | 0.58 | 0.18 |

*Table 7. List of the first 20 subject categories by value of the ratio of highly specialized provinces/active provinces*

| Subject Category | Active provinces | Of which highly specialized | Of which non-specialized | Ratio highly specialized provinces/active provinces | Ratio non-specialized provinces/ active provinces |
|---|---|---|---|---|---|
| Engineering, Petroleum | 17 | 8 | 6 | 0.47 | 0.35 |
| Ornithology | 13 | 6 | 2 | 0.46 | 0.15 |
| Fisheries | 44 | 20 | 16 | 0.45 | 0.36 |
| Agronomy | 51 | 23 | 16 | 0.45 | 0.31 |
| Microscopy | 41 | 17 | 10 | 0.41 | 0.24 |
| Soil Science | 51 | 21 | 19 | 0.41 | 0.37 |
| Emergency Medicine | 32 | 13 | 10 | 0.41 | 0.31 |
| Horticulture | 52 | 21 | 15 | 0.40 | 0.29 |
| Materials Science, Textiles, Paper & Wood | 27 | 10 | 8 | 0.37 | 0.30 |
| Materials Science, Biomaterials | 53 | 19 | 17 | 0.36 | 0.32 |
| Environmental Sciences | 76 | 27 | 22 | 0.36 | 0.29 |
| Agriculture, Multidisciplinary | 62 | 22 | 18 | 0.35 | 0.29 |
| Mining & Mineral Processing | 20 | 7 | 6 | 0.35 | 0.30 |
| Biodiversity Conservation | 45 | 15 | 13 | 0.33 | 0.29 |
| Engineering, Marine | 12 | 4 | 4 | 0.33 | 0.33 |
| Entomology | 48 | 16 | 19 | 0.33 | 0.40 |
| Oceanography | 49 | 16 | 24 | 0.33 | 0.49 |
| Food Science & Technology | 74 | 24 | 25 | 0.32 | 0.34 |
| Hematology | 78 | 25 | 18 | 0.32 | 0.23 |
| Marine & Freshwater Biology | 57 | 18 | 21 | 0.32 | 0.37 |